\documentclass[conference]{IEEEtran}
\IEEEoverridecommandlockouts
\usepackage{cite}
\usepackage{amsmath,amssymb,amsfonts}
\usepackage{algorithmic}
\usepackage{graphicx}
\usepackage{textcomp}
\usepackage{xcolor}
\usepackage{balance}
\usepackage{hyperref}
\usepackage{url}
\ifCLASSOPTIONcompsoc
\usepackage[caption=false,font=normalsize,labelfon
t=sf,textfont=sf]{subfig}
\else
\usepackage[caption=false,font=footnotesize]{subfi
g}
\fi
\def\BibTeX{{\rm B\kern-.05em{\sc i\kern-.025em b}\kern-.08em
    T\kern-.1667em\lower.7ex\hbox{E}\kern-.125emX}}

\begin{document}

\title{A Graph Diffusion Scheme for Decentralized Content Search based on Personalized PageRank\\
}

\author{\IEEEauthorblockN{Nikolaos Giatsoglou}
\IEEEauthorblockA{\textit{ITI - CERTH} \\
Thessaloniki, Greece \\
ngiatsog@iti.gr}

\and

\IEEEauthorblockN{Emmanouil Krasanakis}
\IEEEauthorblockA{\textit{ITI - CERTH} \\
Thessaloniki, Greece \\
maniospas@iti.gr}
\and

\IEEEauthorblockN{Symeon Papadopoulos}
\IEEEauthorblockA{\textit{ITI - CERTH}\\
Thessaloniki, Greece \\
papadop@iti.gr}

\and

\IEEEauthorblockN{Ioannis Kompatsiaris}
\IEEEauthorblockA{\textit{ITI - CERTH} \\
Thessaloniki, Greece \\
ikom@iti.gr}

}

\maketitle

\begin{abstract}
Decentralization is emerging as a key feature of the future Internet. However, effective algorithms for search are missing from state-of-the-art decentralized technologies, such as distributed hash tables and blockchain. This is surprising, since decentralized search has been studied extensively in earlier peer-to-peer (P2P) literature. In this work, we adopt a fresh outlook for decentralized search in P2P networks that is inspired by advancements in dense information retrieval and graph signal processing. In particular, we generate latent representations of P2P nodes based on their stored documents and diffuse them to the rest of the network with graph filters, such as personalized PageRank. We then use the diffused representations to guide search queries towards relevant content. Our preliminary approach is successful in locating relevant documents in nearby nodes but the accuracy declines sharply with the number of stored documents, highlighting the need for more sophisticated techniques.
\end{abstract}

\begin{IEEEkeywords}
decentralized search, nearest neighbors, graph diffusion, Personalized PageRank
\end{IEEEkeywords}

\section{Introduction}
Decentralization is back in the spotlight. While peer-to-peer (P2P) systems were popular in the 2000s, they subsequently lost their appeal to centralized social networks and streaming services. The tide has recently turned for two reasons. The first is the fear over excessive centralization of user data, sparked by pivotal incidents, such as the 2013 leaking of the PRISM surveillance program by Edward Snowden \cite{greenwald2013nsa_prism} and 
the 2016 Cambridge Analytica scandal \cite{cadwalladr2017great_cambridgeanalytica}. The second is the increasing mainstream appeal of cryptocurrencies and blockchain-based applications \cite{chen2018survey_blockchainapps}.

As a result, influential web technologists have called for a decentralized web with actions like the Decentralized Web Summits by the Internet Archives\footnote{\url{https://www.decentralizedweb.net/}} and the Solid project by Tim Berners-Lee
\cite{mansour2016demonstration_solid}. From their part, policy-makers have answered this call, with the European Commission supporting decentralized technologies in its flagship Next Generation Internet initiative\footnote{\url{https://digital-strategy.ec.europa.eu/en/policies/next-generation-internet-initiative}}. 

Currently, the most popular decentralized technologies are distributed hash tables (DHTs) and blockchain. Of these, DHTs come from the previous wave of P2P research and enable document retrieval via unique textual identifiers, with strong guarantees on retrieval delay. For example, the Kademlia DHT powers the Interplanetary File System (IPFS), a decentralized file storage solution that aspires to become a pillar of the decentralized Web\footnote{\url{https://ipfs.io/}}.
On the other hand, blockchain allows decentralized nodes to maintain common states via consensus mechanisms like proof-of-work and broadcasting. Instead of data, blockchain is typically used to broadcast monetary transactions and reward nodes for executing decentralized operations. For example, Filecoin builds on IPFS and uses blockchain to reward nodes for offering file storage. 

But how can one \textit{find} documents in decentralized systems? DHTs require previous knowledge of document identifiers, which must be acquired externally. Alternatively, they can implement distributed inverted indexes by storing relevant document identifiers for search keywords \cite{reynolds2003efficient_dhtsearchengine}, as the  YaCy search engine does\footnote{\url{https://yacy.net/}}.

However, this practice carries fundamental bandwidth and storage constraints\cite{li2003feasibility_dhtwebindexing} and exact keyword matching is dated compared to the semantic awareness of modern search engines. On the other hand, unstructured search techniques, such as flooding, random walks, index sharing, and query caching\cite{khatibi2021resource_unstructuredsurvey} often suffer from high communication overhead and unpredictable delays. Finally, blockchain has been used to reward nodes for executing indexing and retrieval operations in decentralized search engines, such as Presearch\footnote{\url{https://presearch.org/}}, but broadcasting indexes to all nodes is prohibitive in terms of bandwidth and storage.

While research on decentralized search has stagnated on the above bottlenecks, centralized search engines have evolved to better understand query semantics. This evolution has been driven by advancements in \textit{embeddings}, latent representations of text and other types of content \cite{lin2021pretrained_transformersforretrieval}. Retrieval with embeddings often follows a vector space model, which extracts vector representations for documents and queries and compares their relevance with a simple similarity metric, such as the dot product or cosine similarity. This way, the retrieval can be cast as a \textit{nearest-neighbor} problem, which tries to find the nearest documents to a query according to the selected similarity metric. In contrast to term-frequency vectors, embeddings are lower-dimensional and enable semantic rather than exact term matching, giving rise to \textit{dense retrieval}.

Here, we argue that decentralized search can benefit from modern techniques employed by centralized search engines. To this end, we revisit the decentralized search problem from an embedding-based standpoint. We further employ a graph signal processing technique to implement similarity-based P2P query routing. We propose composing node embeddings from local node documents and diffusing them through P2P networks with decentralized implementations of graph filters, such as Personalized PageRank (PPR). We then use the diffused embeddings to guide decentralized search towards nodes with relevant documents. We experiment with a simulation of a real-world P2P network and investigate how our solution scales with the number of documents in the network. Our approach successfully locates relevant documents in nearby nodes but accuracy sharply declines as the number of documents increases, highlighting the need for further research.

\section{Background and Related Work}
This section explores related work on decentralized search (Subsection~\ref{sec:decentralized search}) and then presents dense information retrieval (Subsection~\ref{sec:dense retrieval}) and graph signal processing (Subsection~\ref{sec:gsp}) background to contextualize later analysis.

\label{sec:background}
\subsection{Decentralized search} \label{sec:decentralized search}
Decentralized search received attention in the early 2000s for P2P file sharing systems, such as Gnutella and Freenet \cite{aberer2002overview_overviewgnutellaetal}. Gnutella introduced \textit{flooding}, the simplest technique for search, which forwards search queries to all nodes within a specified number of hops. As P2P platforms grew in size, flooding was soon found to not scale in terms of bandwidth consumption \cite{ritter2001gnutella_gnutellascalability}, giving rise to alternatives, such as random walks, index sharing, and super-peer architectures \cite{lua2005survey_earlysurveyonp2p}. Of these, \textit{informed} methods exploit hints about possible document locations and outperform \textit{blind} methods, like flooding and random walks in terms of delay and communication cost. This comes at the expense of costly state maintenance at nodes \cite{tsoumakos2006analysis_blindinformeddistinction}. 
\par
Informed search methods rely on query routing and can be further categorized into \textit{document-} and \textit{query-oriented} ones \cite{arour2015learning_querycontentoriented}. In document-oriented methods, P2P nodes exchange information about their stored documents \cite{crespo2002routing_firstdocumentrouting,  kumar2005efficient_bloomfilters}. As the storage cost increases with the number of documents in the network, the advertisement radius is limited and summarization is employed to compress the advertisements, for instance with Bloom filters \cite{kumar2005efficient_bloomfilters}. Both techniques introduce routing errors. In query-oriented methods, nodes store information of passing queries and their results \cite{kalogeraki2002local_firstqueryrouting, li2006improve_queryroutingwithrl} and, when a new query arrives, it is forwarded to the most successful route travelled by similar past queries. These methods are attractive because they avoid storing information about unpopular documents. On the other hand, they are blind to unseen queries, especially at the beginning of the network's operation when no information is available (cold-start problem).

While informed search identifies the locations of relevant documents through routing, DHTs decouple these two operations with a clever application of hashing \cite{lua2005survey_earlysurveyonp2p}. In particular, DHT nodes agree to store documents whose hash values are the closest to their own address, according to a distance function. As a result, when nodes search for a document, they can resolve its location and reach it through routing. For efficiency, most DHT systems, such as Chord, Pastry, and Kademlia, structure P2P networks so that all locations are reachable within a maximum number of hops \cite{lua2005survey_earlysurveyonp2p}, although this structuring is not strictly required\footnote{Efficient addressing can be enforced on networks with arbitrary structure, for example with greedy embeddings \cite{hofer2013greedy_greedyembeddings}.}.

The theoretical properties and practicality of DHTs have made them attractive for modern decentralized systems, such as IPFS, but they are best suited for key-based retrieval. For other types of search, such as range and nearest neighbor queries, adaptations or other distributed data structures are needed, such as skip-lists and skip-graphs  \cite{reynolds2003efficient_dhtsearchengine, bongers2015survey_multidimensionalrangequeries, gao2007efficient_dhtssimilaritysearch}. These solutions carry their own limitations, including security concerns and poor load balancing of traffic.

\subsection{Dense retrieval}\label{sec:dense retrieval}
Information retrieval is often based on vector space models that represent documents and queries as vectors and estimate document relevance to queries via a similarity metric. Text vector representations are traditionally derived from bag-of-words models based on word frequencies, predominantly the TF-IDF and BM25 models\cite{manning2010introduction}. Those yield high-dimensional sparse vectors that can be efficiently stored in inverted index tables but do not capture the underlying semantics, such as implied contexts, synonyms, or word co-usage patterns. To address this issue, research has moved towards lower-dimensional dense representations, which encode latent semantics and enable soft matches. Dense retrieval has recently demonstrated definite improvement over sparse retrieval (represented by the BM25 model) \cite{lin2019neural_neuralhype}, owing to the successful transfer of deep learning advances \cite{lin2021neural_neuralhyperecant}. 

Key steps in this process have been the development of efficient vector representations for words with the Word2Vec and Glove frameworks \cite{pennington2014glove_glove}, which were later extended to sentences. While sentence embeddings are less understood, they were shown to capture linguistic information \cite{conneau2018you_sentenceembeddings} and are useful to retrieval \cite{yang2019simple_sbertforadhoc}. 

Currently, the state of the art for dense retrieval focuses on pre-trained transformer models, commonly based on BERT \cite{devlin2018bert}, 
which are subsequently fine-tuned on downstream retrieval tasks \cite{lin2021pretrained_transformersforretrieval}. There are two extreme approaches in using BERT for retrieval, \textit{cross-encoders} and \textit{bi-encoders}. Cross-encoders consider all interactions among query and document words, which yields the best accuracy but with high processing and energy costs. For instance, cross-encoders need to process all documents and queries at query-time, which incurs unreasonable delays. In contrast, bi-encoders conform to the vector space model in that documents and queries are transformed separately to vectors and interact via simple operations, such as the dot product or cosine similarity. While bi-encoders are less accurate than cross-encoders, they outperform BM25, enable proactive document indexing, and their inference is quick and cheap with approximate nearest-neighbor algorithms \cite{aumuller2020ann}. Therefore, the vector space model and nearest-neighbor algorithms remain relevant for modern search applications.

\subsection{Graph signal processing}\label{sec:gsp}
Graph signal processing is a recently popularized field that generalizes traditional signal processing principles to graphs \cite{ortega2018graph_gspsurvey,huang2018graph}. With this approach, graph signals are defined as collections of node values, e.g., scalars, vectors, and graph filters study their propagation through graphs. In particular, a graph convolution operation is defined, which performs one-hop propagation of node values through matrix multiplication, and graph filters are defined by weighted aggregation of multihop propagations. Popular graph filters, such as PPR and heat kernels perform the equivalent of low-pass filtering by placing higher importance to node values that are propagated fewer hops away.
\par
When node values are vectors, graph filters operate independently on each vector dimension. This type of propagation is useful by itself for downstream predictive tasks, such as prediction propagation in graph neural networks \cite{klicpera2018predict,dong2021equivalence}. In this work, we consider low-pass graph filters as a type of smoothing that concentrates around a small area around nodes. This area can be tuned by a single parameter of the PPR filter.

\section{Problem Setting}
\label{sec:problem_seting}
This section first presents dense retrieval operations, as they would be applied by modern centralized search engines (Subsection \ref{subsec:centralized_setting}), and then re-formulates them in a decentralized setup (Subsection \ref{subsec:decentralized_setting}).

\subsection{Centralized Setting}
\label{subsec:centralized_setting}

In the centralized setting, we consider search engines that are responsible for answering queries over collections of stored documents $\mathcal{D}$. When engines receive queries $q$, they compute relevance scores $s(d,q)$ for all documents $d \in \mathcal{D}$. They then estimate the top-$k$ most relevant documents per
\begin{equation}
    \underset{d \in \mathcal{D}}{\text{arg top-}k} ~ s(d, q).
\end{equation}

In this paper, we consider the bi-encoder model of dense retrieval, which splits the score computation in two parts: i) an \textit{encoding} part that transforms queries $q$ and documents $d$ to $\nu$-dimensional embedding vectors $\mathbf{e}_q, \mathbf{e}_d$ respectively ($\mathbf{e}_q, \mathbf{e}_d \in \mathbb{R}^\nu$), and ii) a \textit{comparison} part that derives the score $s$ from the embeddings. This is formalized as 
\begin{equation}
\label{candidate_document}
    s = \phi(\mathbf{e}_q, \mathbf{e}_d) = \phi\left(\eta_q (q), \eta_d (d)\right)
\end{equation}
where $\eta_q, \eta_d$ are encoding functions for queries and documents respectively, and $\phi$ is a comparison mechanism \cite{deepretrievalframework}. The above formulation is attractive because it contains the computational complexity to the encoding function $\eta$, which can be pre-computed during indexing. In contrast, the comparison function $\phi$ is executed at query time and is therefore chosen to be computationally lightweight; usually, the dot product or cosine similarity is chosen\footnote{These are equivalent when the embeddings are L2-normalized.}. These choices cast the retrieval as a $k$ nearest-neighbor problem, which can be computed efficiently with popular approximation algorithms, e.g., based on locality sensitive hashing or hierarchical navigable small world graphs \cite{aumuller2020ann}. 

\subsection{Decentralized Setting}
\label{subsec:decentralized_setting}

To move to the decentralized setting, we consider a P2P network whose nodes maintain their own private document collections. The network is modeled as an \textit{undirected} graph $\mathcal{G}=(\mathcal{V}, \mathcal{E})$, where $\mathcal{V}$ is the set of nodes and $\mathcal{E}\subseteq \mathcal{V}\times \mathcal{V}$ their communication edges, while  $\mathcal{D}_u\subseteq\mathcal{D}$ represents the local documents of node $u$.
\par
When nodes initiate queries, they first execute the retrieval operations of subsection \ref{subsec:centralized_setting} over their local document collections, and then forward queries to their one-hop neighbors to retrieve more results. Farther nodes can be contacted by relaying the queries along nodes. Since contacting all nodes would induce non-scalable communication costs and delays, we allow the search to fail to find relevant documents, even if these could have been retrieved by centralized search engines. The goal of our analysis is to make clever forwarding decisions to achieve high search hit accuracy of relevant documents.

\section{Diffusion-based decentralized search}
\label{sec:proposal}
Our decentralized scheme for search is a document-oriented solution where nodes maintain a summary of documents available from their neighbors. These summaries take the form of \textit{node embedding} vectors, denoted by $\mathbf{e}_u$, which are composed from the embeddings of both local and nearby documents. To generate the node embeddings, when new nodes enter the network or update their document collections, they compute \textit{personalization vectors}, denoted by $\mathbf{e}_u^{(0)}$, which characterize their local document collections (Subsection \ref{subsec:personalization}). Subsequently, the nodes diffuse their personalization vectors to the network with an iterative and asynchronous diffusion algorithm based on PPR (Subsection \ref{subsec:diffusion}). This algorithm converges to the node embedding vectors and also keeps track the embeddings of the one-hop neighbors for each node. At query-time, the nodes can use their stored neighbor embeddings to forward queries towards promising next hops (Subsection \ref{subsec:forwarding}).

\subsection{Node personalization}
\label{subsec:personalization}
 Ideally, for each node $u$, we would like to estimate the maximum score of all neighbors $v$, as in \eqref{candidate_document}, without knowing their documents $\mathcal{D}_v$. A simple way is to represent each node with the personalization vector $\mathbf{e}_u^{(0)}$ that is the sum of the node's document embeddings. This has the attractive property that,  due to the linearity of the interaction function, the dot product of the query with the neighbor embedding yields the total relevance of the neighbor's documents:
\begin{equation}
    \mathbf{e}_q \cdot \mathbf{e}_v^{(0)} = \mathbf{e}_q \cdot \sum_{d \in \mathcal{D}_v} \mathbf{e}_d= \sum_{d \in \mathcal{D}_v} \mathbf{e}_q \cdot \mathbf{e}_d.
\end{equation}
This approach tends to score higher nodes with a larger number of documents. This is desirable in general although it runs the risk of prioritizing nodes with many irrelevant documents over nodes with a few but relevant documents. 

\subsection{Diffusion of embeddings}
\label{subsec:diffusion}
After computing their personalization vectors, the nodes transmit them to their neighbors. Instead of traditional $n$-hop advertising, we consider a diffusion scheme based on graph signal processing. A typical diffusion has the form:
\begin{equation}
\label{diffusion}
\mathbf{E} = \mathbf{H} \mathbf{E}^{(0)} ~ \Rightarrow ~
    \mathbf{e}_u = \sum_{v \in V} h_{u v} \mathbf{e}_v^{(0)}
\end{equation}
where $\mathbf{E}^{(0)}$, $\mathbf{E}$ are the initial and diffused embeddings in matrix form, $\mathbf{H}$ is the weight matrix or impulse response of diffusion, whose elements $h_{u v}$ represent the impact of node $v$ to $u$. While the diffusion weights $\mathbf{H}$ could be learned with a machine learning algorithm, the complexity of learning would scale with $\mathcal{O}(N^2)$, which would be intractable for large graphs. Therefore, we have chosen the PPR algorithm for calculating the weights, which is a popular approach in the literature \cite{klicpera2018predict}, and can be implemented in a decentralized and asynchronous way \cite{krasanakis2021p2pgnn_asynchronousppr}, which is a highly desirable feature. 

In PPR, we associate $h_{u v}$ with the probability to reach $v$ via a random walk that starts from $u$. If the random walk were allowed to progress, as in the traditional PageRank, it would forget its origin $u$ and converge to a probability characterizing only $v$. To avoid this, in PPR, we force the walker to teleport back to node $u$ with probability $a$. Thus,  $h_{u v}$ is associated with the probability to reach node $v$ from $u$ with a short walk of average length $1/a$.

Formally, denoting by $\boldsymbol{\pi} [v]$ the probability of arriving at node $v$, and by $\boldsymbol{\delta}_u[v]$ the one-hot vector at node $u$, i.e., $\boldsymbol{\delta}_u[u]=1$ and $\boldsymbol{\delta}_u[v]=0$ for $v \neq u$, we have
\begin{equation}
\label{ppr_recursive}
    \boldsymbol{\pi} [v] = (1-a) \mathbf{A} \boldsymbol{\pi} [v] + a \boldsymbol{\delta}_u[v] \Rightarrow \boldsymbol{\pi} [v] = a (\mathbf{I}-(1-a)\mathbf{A})^{-1} \boldsymbol{\delta}_u[v]
\end{equation}
where $\mathbf{I}$ is the identity matrix and $\mathbf{A}$ the transition matrix of the Markov chain, based on a suitable normalization of the adjacency matrix of $\mathcal{G}$ or external weights. Considering the definition of $\boldsymbol{\delta}_u[v]$, it is clear that the columns of $a (\mathbf{I}-(1-a)\mathbf{A})^{-1}$ correspond to the desired probabilities for different origins $u$. The diffused embeddings of \eqref{diffusion} are thus given by:
\begin{equation}
\label{ppr_solution}
\mathbf{E} = a (\mathbf{I}-(1-a)\mathbf{A})^{-1} \mathbf{E}^{(0)}
\end{equation}
While the embeddings are propagated to the whole graph, the \textit{effective} range of the diffusion is tuned by the parameter $a$. 

For the decentralized and asynchronous implementation, we first express \eqref{ppr_solution} iteratively as:
\begin{equation}
\label{ppr_iterative}
    \mathbf{E}^{(t)}  = (1-a) \mathbf{A} \mathbf{E}^{(t-1)} + a \mathbf{E}^{(0)},
\end{equation}
which converges to \eqref{ppr_solution} but is synchronous. Subsequently, we make the iteration \textit{asynchronous} by letting node pairs exchange and update embeddings. As proven in \cite{krasanakis2021p2pgnn_asynchronousppr}, if the update intervals are not arbitrarily long, the embeddings converge to \eqref{ppr_recursive} in distribution, which is a good approximation of the centralized scheme.

\begin{figure}[!t]
\centering
\includegraphics[width=2.5in]{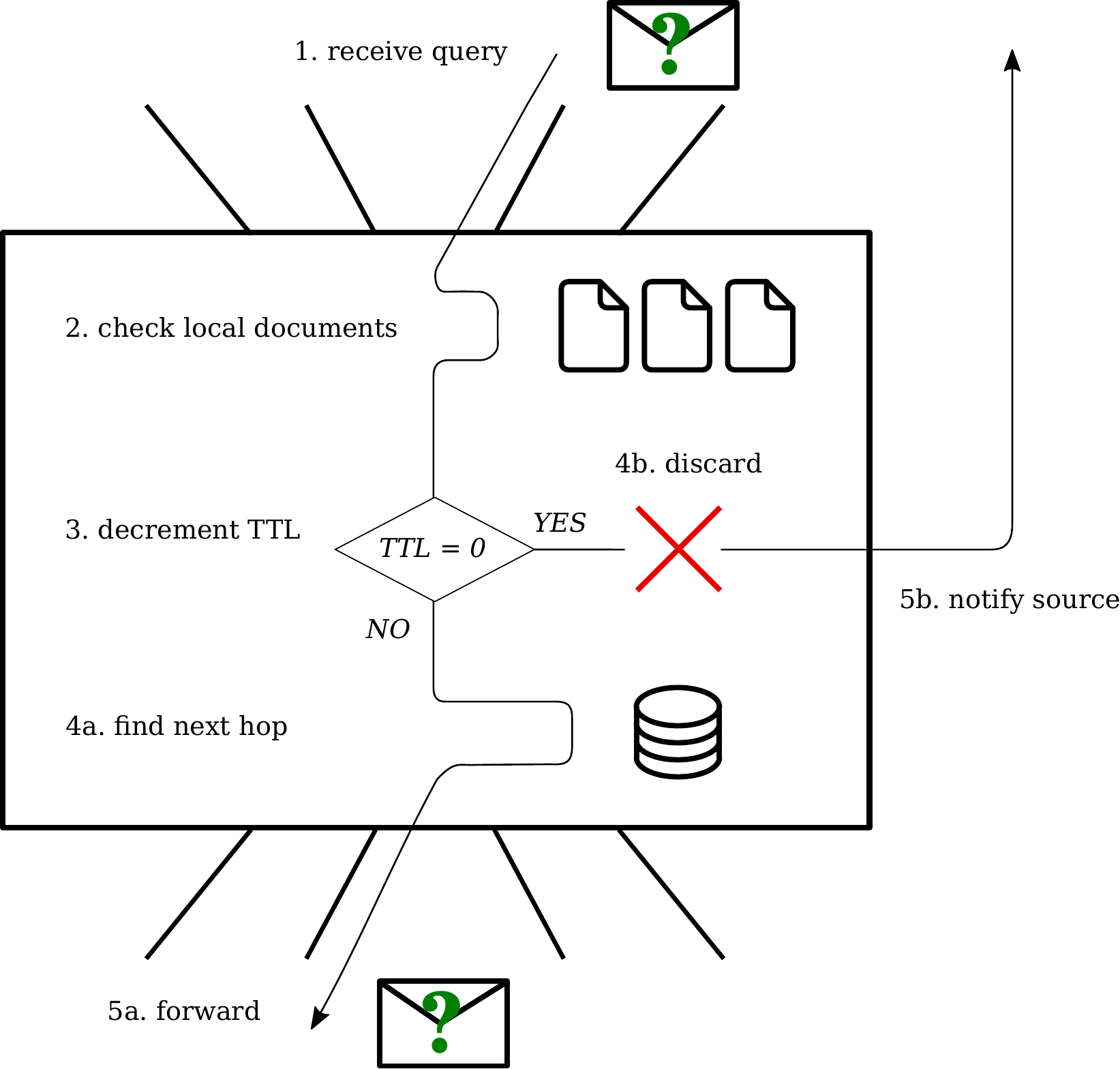}
\caption{Node operations when a query is received.}
\label{img:node_operations}
\end{figure}

\subsection{Forwarding operations}
\label{subsec:forwarding}
Node embeddings are used at query-time to guide search towards promising nodes, essentially performing a biased random walk. Queries keep track of the $k$ most relevant documents they have encountered along with their relevance score\footnote{If documents are too large, the message can track the IP addresses of the source nodes or content identifiers if available, e.g., IPFS content IDs.}. Since visiting all nodes in the network is impractical, we impose a maximum number of hops with a time-to-live (TTL) field in the query message, which helps prevent queries from circulating in the network indefinitely. Due to the TTL limitation, we prioritize unvisited nodes for forwarding. To this end, the nodes keep track of the neighbors from which they have received and to which they have sent messages. We purposefully reject the alternative (and slightly more efficient) solution of recording the visited nodes in the query message in order to protect the privacy of node connections. In our solution, nodes relay the queries recursively, i.e., from node to node, and when their TTL expires, a response message is returned to the querying nodes via backtracking.

Fig. \ref{img:node_operations} illustrates the node operations when a new query arrives. As described in subsection \ref{subsec:decentralized_setting}, nodes first evaluate the query on their local documents according to the retrieval operations of subsection \ref{subsec:centralized_setting}. Afterwards, they decrement the TTL field of the query message by 1 and check if the message is still alive. If the TTL has expired, the nodes discard the query and send a query response message to the reverse path, otherwise, they commence the forwarding procedure: nodes first determine a set of candidate next hops from their neighbors, which excludes previously visited nodes remembered by the nodes\footnote{If no neighbors remain after this step, nodes consider all their neighbors as candidates as we do not want to waste opportunities for forwarding considering the TTL limitation.}. Nodes then match via dot product the embeddings of the candidate next hops with the query embedding, and select a few neighbors with the highest score. When a single neighbor is selected, the outcome is a simple random walk, otherwise, multiple walks are executed in parallel.

\section{Experimental Evaluation}\label{sec:experiments}
\label{sec:setup}
We evaluate a retrieval operation in a social P2P network based on two datasets: a social network graph and a corpus of pre-trained embeddings (Subsection~\ref{sec:datasets}). Through simulation (Subsection~\ref{sec:simulation}), we investigate the scalability of our scheme with the number of stored documents in the network, $M$, in terms of the hit accuracy (Subsection~\ref{sec:acc}) and the average number of hops of successful queries (Subsection~\ref{sec:hop}).

\subsection{Datasets}
\label{sec:datasets}
Experiments are conducted on the Facebook social circles graph\cite{leskovec2012learning_fbdataset} hosted by the SNAP project\footnote{\url{http://snap.stanford.edu}}. This is an undirected graph of 4,039 Facebook users (nodes) and their 88,234 friend relations (edges). We consider this graph representative of P2P networks built on top of social relations, which are expected to resemble friend relations of centralized social networks.

Documents and queries are represented using 300-d word embeddings, trained by the Glove model on Wikipedia articles \cite{pennington2014glove_glove} and distributed by the GenSim library\footnote{\url{https://radimrehurek.com/gensim/}}. While Glove embeddings are not ideal for retrieval, they are good predictors of similarity with the cosine similarity metric. As mentioned in Section \ref{sec:setup}, the nearest-neighbor search mechanism is independent from the embedding method, which allows us to study search in isolation. In fact, queries and documents can refer to any type of content, even multimedia, provided relevance is a linear function of their embeddings.

\begin{figure}[!t]
   \centering
\fbox{\parbox{0.8\linewidth}{\begin{algorithmic}[1]
    \STATE Generate documents and queries from Glove
    \STATE Distribute $N$ documents uniformly over $\mathcal{G}$
    \STATE Compute node embeddings
    \REPEAT
    \STATE Diffuse node embeddings asynchronously
    \UNTIL embeddings converge
    \STATE Distribute queries
    \REPEAT
    \STATE Forward queries
    \UNTIL all queries expire
\end{algorithmic}}}
\caption{Pseudo-code for the simulation of the decentralized search setting.}
\label{fig:pseudocode}
\end{figure}

\subsection{Simulation setup}
\label{sec:simulation}
Fig. \ref{fig:pseudocode} presents our simulation in pseudo-code. We first generate queries and documents from the Glove dataset using 1000 random words as queries and their nearest neighbors as gold documents, provided that their cosine similarity is over 0.6 and the two sets do not overlap. The remaining words are treated as a pool of irrelevant documents. We further distribute the documents over the graph's nodes uniformly and compute the node embeddings. This is followed by a warm up period, in which we diffuse the node embeddings over the network with the asynchronous PPR algorithm. The algorithm runs until the embeddings converge.

We then proceed with evaluating the top-$1$ document retrieval performance over sampled queries, whose number depends on the simulation scenario. In each iteration, queries are distributed over the network and are forwarded independently. For simplicity, each query performs a simple random walk, which is the most challenging case and can be easily extended to parallel walks. In the future, we plan to investigate parallel walks more thoroughly along with time-evolving conditions and the top-$k$ performance. More realistic document distributions are also worthwhile; in fact, they are expected to aid diffusion, since they naturally exhibit spatial correlation.

\subsection{Hit Accuracy}
\label{sec:acc}
\begin{figure*}[!t]
\centering
\subfloat[]{\includegraphics[width=0.41\linewidth]{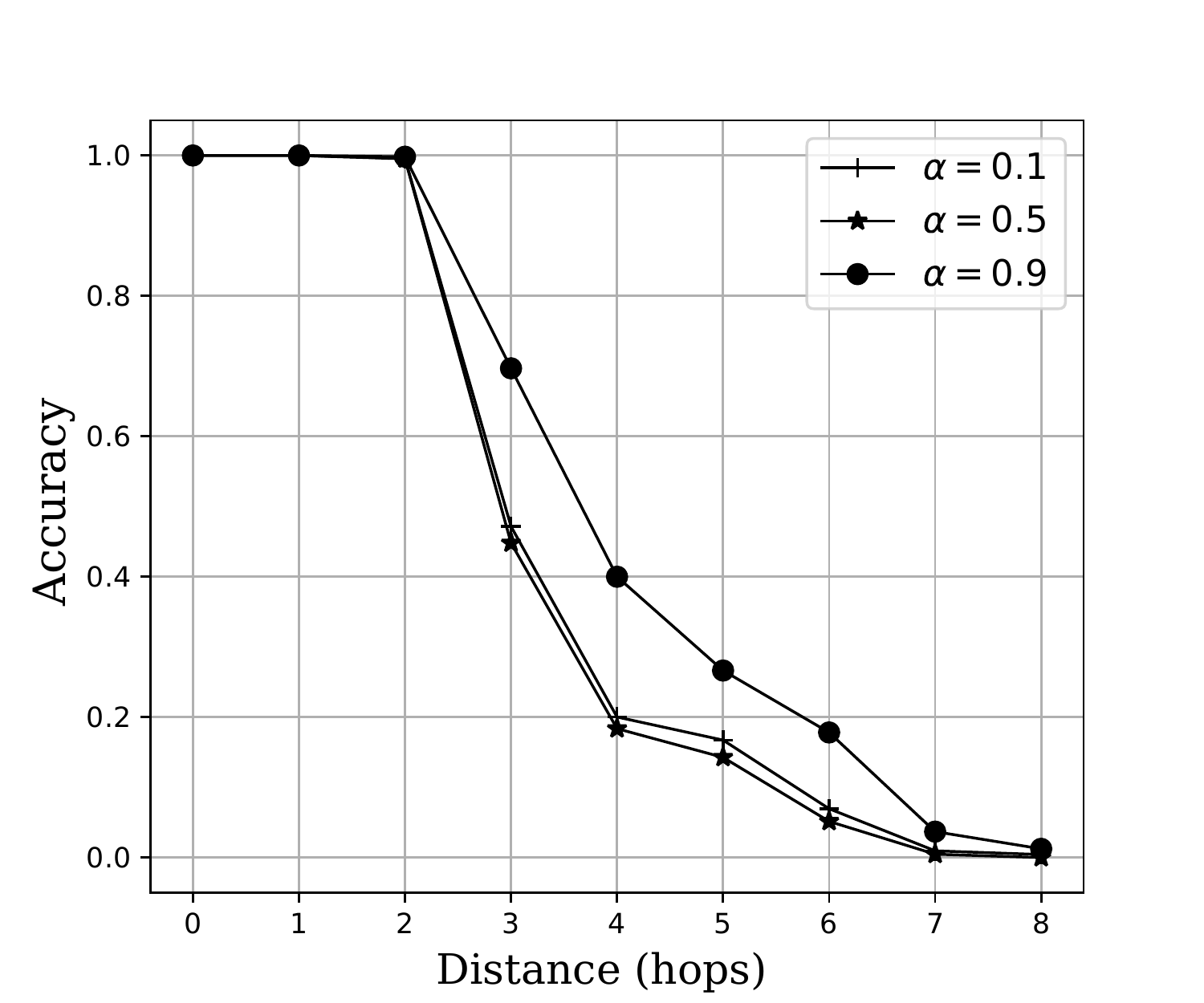}
\label{fig:10docs}}
\subfloat[]{\includegraphics[width=0.41\linewidth]{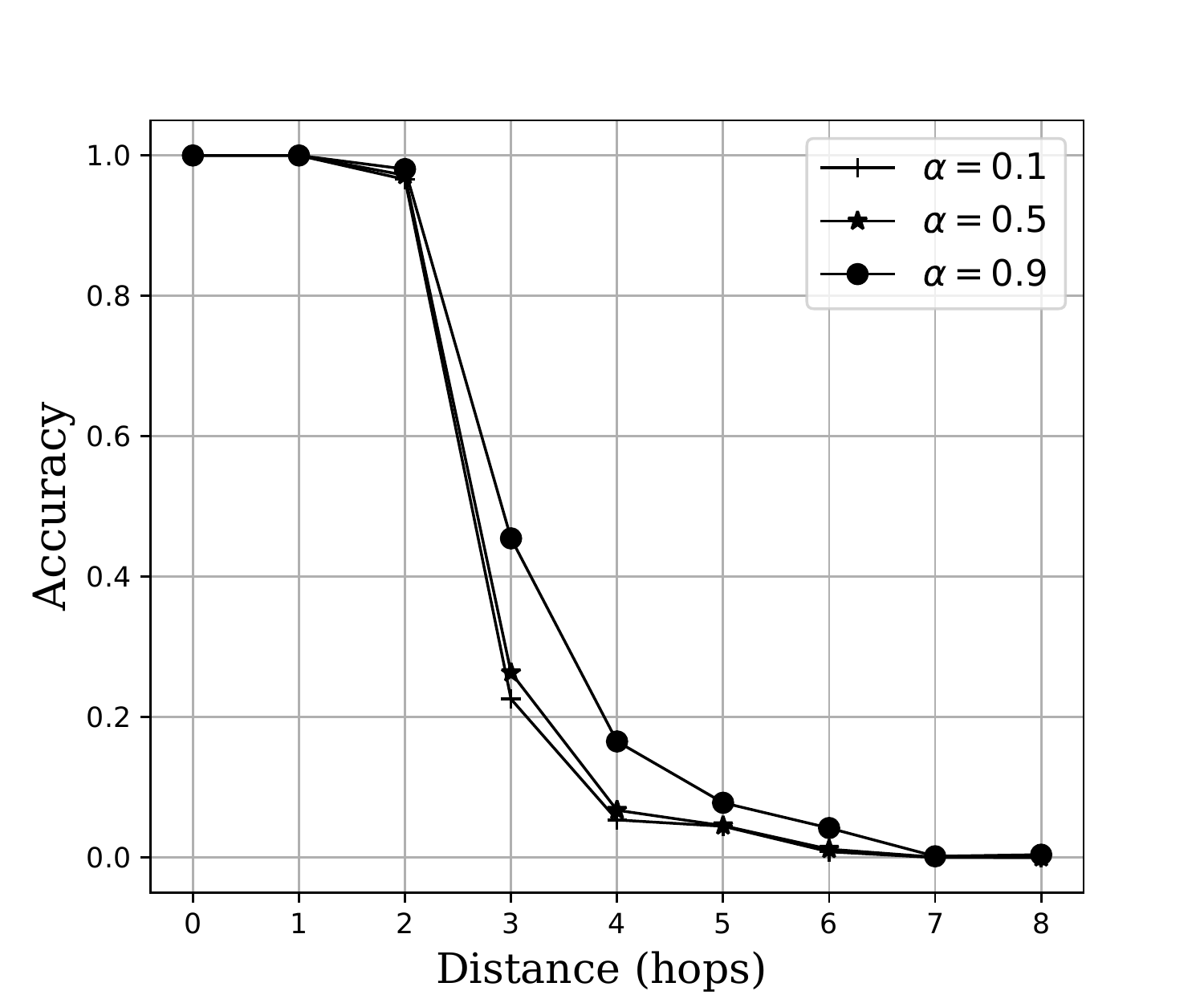}
\label{fig:100docs}}

\subfloat[]{\includegraphics[width=0.41\linewidth]{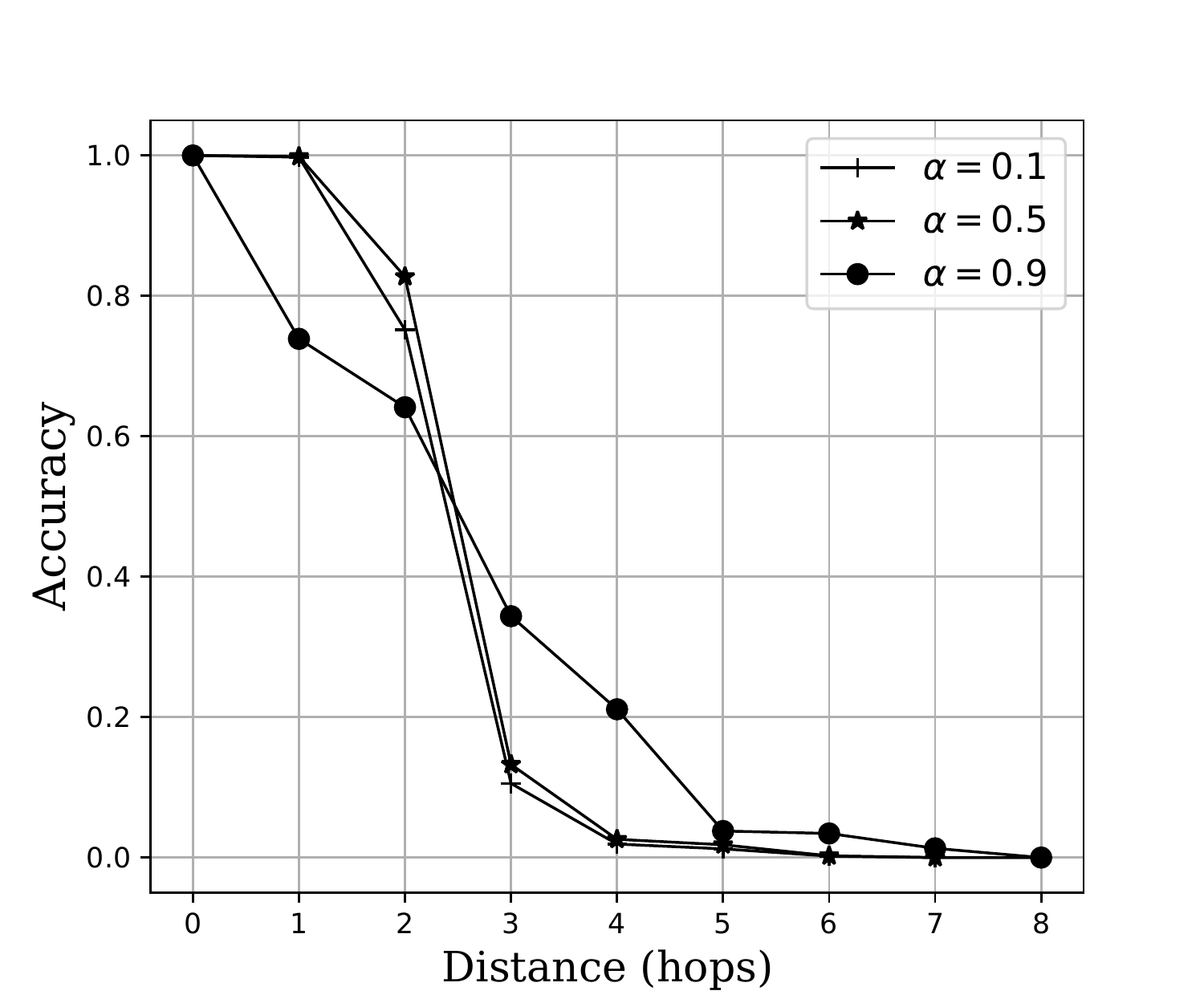}
\label{fig:1000docs}}
\subfloat[]{\includegraphics[width=0.41\linewidth]{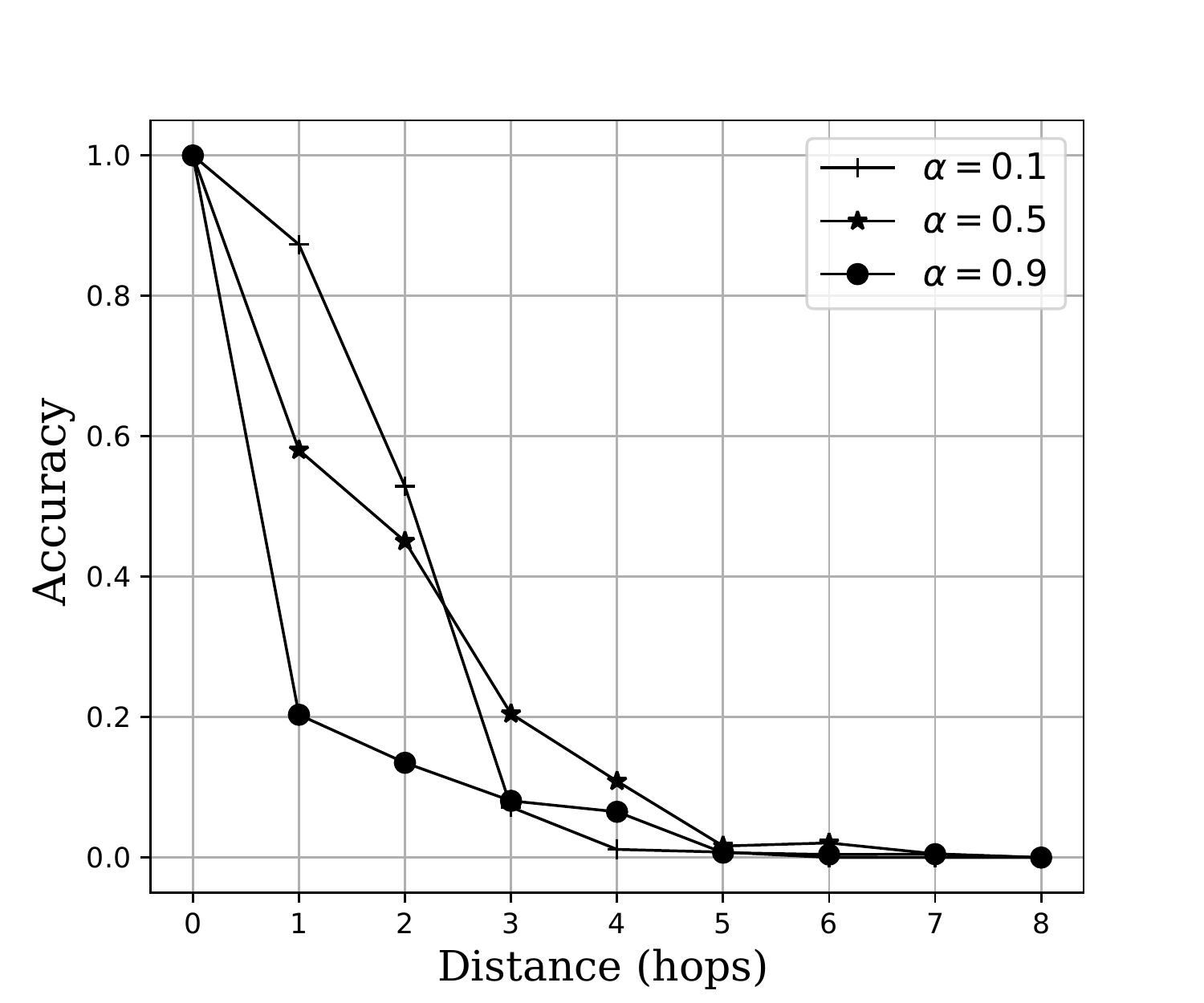}
\label{fig:10000docs}}
\caption{Accuracy analysis for a) 10, b) 100, c) 1000, and d) 10000 documents in the network.}
\label{fig:acc_analysis}
\end{figure*}

In this series of experiments, we evaluate the accuracy of our algorithm over the number of stored documents in the network, $M$, and the teleport probability of PPR, $\alpha$, which determines the average diffusion radius. For $M$, we select $10$, $100$, $1000$, and $10000$ documents to investigate 4 orders of magnitude. In each iteration, we store one gold and $M$-1 irrelevant documents in the network, and sample multiple querying nodes, one from each radius away from the location of the gold document. At the end of simulation, the accuracy is computed as the percentage of queries that retrieved the gold document within a TTL of 50 hops. The simulation is repeated for three different values of $\alpha$, $0.1$, $0.5$, and $0.9$, as examples of heavy, moderate, and light diffusion respectively. The results are depicted in Fig. \ref{fig:acc_analysis}.

Figs. \ref{fig:10docs} and \ref{fig:100docs} show that our algorithm excels at finding documents within 2 hops away, provided that there are few documents in the network. In contrast, the accuracy starts to decline at 3 hops and deteriorates significantly farther away. Surprisingly, heavy diffusion does not aid accuracy, as more documents are discovered when the teleport probability is 0.9. The results change radically with more stored documents. In Figs. \ref{fig:1000docs} and \ref{fig:10000docs}, we see that the accuracy remains high mainly for documents in neighboring nodes and the impact of $\alpha$ is more varied. In this case, heavier diffusion is better at small distances although $a=0.9$ appears beneficial at 3 and 4 hops when the stored documents are 1000. With 10000 documents, the performance deteriorates considerably.

The above show that the PPR diffusion is useful for local neighborhood search but its accuracy declines with the number of stored documents. This is attributed to the loss of information for individual documents when many embeddings are summed, either during summarization or diffusion. The behavior with $\alpha$ can also be explained by the following trade-off: heavy diffusion (low $\alpha$) announces documents within a wider range but adds more noise due to the summation of the embeddings. In contrast, light diffusion (high $\alpha$) adds less noise but may fail to notify nearby nodes. Considering this trade-off, when few documents are stored in the network (Figs. \ref{fig:10docs} and \ref{fig:100docs}), it is preferable to leave fewer and cleaner hints as the random walk will eventually find the correct document. In contrast, with more documents (Figs. \ref{fig:10docs} and \ref{fig:100docs}), there is already noise in the network and light diffusion may hinder the random walk from finding documents even 1 hop away. 

\subsection{Hop Count Analysis}
\label{sec:hop}
In this experiment, we compute the average hop count for successful queries until the gold document is found. As in Section \ref{sec:acc}, the queries are considered successful when they retrieve the correct document within 50 hops. We note that, since the queries do not know when they find the gold document and must complete their TTL, the average hop count does not indicate bandwidth consumption but can guide the choice of TTL. For the setup, we execute 500 iterations in each of which we distribute 10 queries uniformly in the network, for a total of 5000 samples. We also choose the value 0.5 for the teleport probability $\alpha$, scale the number of documents for 10 to 10000, and randomize the document distribution at each iteration, as in the accuracy experiment. Our results are summarized in Table \ref{hop_analysis}.

Table \ref{hop_analysis} shows that less queries are successful when the stored documents increase, consistently with the accuracy results of Section \ref{sec:acc}. Furthermore, with more documents, longer walks are required as both the median and the mean hops to reach the gold documents increase. The discrepancy between the median and the mean hops implies a skewed distribution, i.e., a few walks succeed after a large number of hops and drive the mean higher, which is corroborated by the high standard deviation. Combined with the results of the accuracy experiment, the above show that, even though documents are found predominantly by nearby nodes, some queries need to circulate for additional hops until they succeed. It is encouraging though that success is still possible with a high number of documents, such as 10000.

\begin{table}[!t]
\renewcommand{\arraystretch}{1.3}
\caption{Average Hop Count}
\label{hop_analysis}
\centering
\begin{tabular}{c|c|c|c|c}
\hline
$M$ documents & success rate & median hops & mean hops & std hops \\
\hline
10 & 1905 / 5000 & 3 & 7.62 & 10.83 \\
100 & 1265 / 5000& 4 & 11.21 & 13.37 \\
1000 & 1054 / 5000 & 9 & 15.26 & 14.55 \\
10000 & 877 / 5000 & 9 & 14.31 & 13.36 \\
\hline
\end{tabular}
\end{table}

\section{Conclusions}
\label{sec:conclusions}
As decentralization is becoming an increasingly important feature of the future Internet, new algorithms are needed for effective decentralized search. In this paper, we revisit this long-standing problem from a combined embedding and graph diffusion perspective. Specifically, considering a P2P network with nodes of only local knowledge over their document collections, we apply the PPR algorithm to diffuse summarized information about the documents in the network. Our results show that this diffusion can be beneficial for local neighborhood search but further enhancements are needed to improve the performance for global search. Our current line of research is to exploit correlations in the document distribution and derive more sophisticated aggregation methods that encode more information about the grouped documents. 

\section*{Acknowledgment}
This research was supported by the EU H2020 projects AI4Media (Grant Agreement 951911), MediaVerse (GA 957252) and HELIOS (GA 825585). 
The authors want to thank Dr. Ioannis Sarafis for his productive feedback on the decentralized search scheme.

\balance
\bibliographystyle{IEEEtran}
\bibliography{IEEEabrv, dinps2022}

\end{document}